%% file: main.tex
\documentclass[reprint,prx,aps,amsmath,amssymb,floatfix,superscriptaddress]{revtex4-2}
\usepackage{physics}
\usepackage{dcolumn}
\usepackage{bm}
\usepackage{graphicx}
\usepackage{amsthm}
\usepackage{color}
\usepackage{physics}
\usepackage{xcolor}
\usepackage{verbatim}
\usepackage{multirow}
\usepackage{esint}
\usepackage[english]{babel}
\usepackage{amsfonts}
\usepackage{slashed}
\usepackage{latexsym}
\usepackage{bm}
\usepackage[colorlinks = true,
            linkcolor = red,
            urlcolor  = blue,
            citecolor = magenta,
            anchorcolor = red]{hyperref}
\usepackage{esint}
\usepackage{soul}
\usepackage{cancel}
\usepackage[normalem]{ulem}
\usepackage{empheq}
\usepackage{dsfont}
\usepackage{amsmath}
\usepackage{ulem}
\usepackage{epsfig}
\usepackage{enumerate}
\definecolor{sapphire}{rgb}{0.03, 0.03, 0.41}
\DeclareMathAlphabet\mathbfcal{OMS}{cmsy}{b}{n}
\def\be{\begin{equation}}
\def\ee{\end{equation}}

\begin{document}

\title{ High-Coherence Kerr-cat qubit in 2D architecture}



\author{Ahmed Hajr}
\email{These authors contributed equally; Correspondence should be addressed to: hajr@Berkeley.edu}
\affiliation{Quantum Nanoelectronics Laboratory, Department of Physics, University of California at Berkeley, Berkeley, California 94720, USA}

\author{Bingcheng Qing}
\email{These authors contributed equally; Correspondence should be addressed to: hajr@Berkeley.edu}

\author{Ke Wang}
\affiliation{Quantum Nanoelectronics Laboratory, Department of Physics, University of California at Berkeley, Berkeley, California 94720, USA}

\author{Gerwin Koolstra}
\affiliation{Quantum Nanoelectronics Laboratory, Department of Physics, University of California at Berkeley, Berkeley, California 94720, USA}

\author{Zahra Pedramrazi}
\affiliation{Quantum Nanoelectronics Laboratory, Department of Physics, University of California at Berkeley, Berkeley, California 94720, USA}
\affiliation{Computational Research Division, Lawrence Berkeley National Laboratory, Berkeley, California 94720, USA}

\author{Ziqi Kang}
\affiliation{Quantum Nanoelectronics Laboratory, Department of Physics, University of California at Berkeley, Berkeley, California 94720, USA}

\author{Larry Chen}
\affiliation{Quantum Nanoelectronics Laboratory, Department of Physics, University of California at Berkeley, Berkeley, California 94720, USA}

\author{Long B. Nguyen}
\affiliation{Quantum Nanoelectronics Laboratory, Department of Physics, University of California at Berkeley, Berkeley, California 94720, USA}

\author{Christian Jünger}
\affiliation{Quantum Nanoelectronics Laboratory, Department of Physics, University of California at Berkeley, Berkeley, California 94720, USA}
\affiliation{Computational Research Division, Lawrence Berkeley National Laboratory, Berkeley, California 94720, USA}

\author{Noah Goss}
\affiliation{Quantum Nanoelectronics Laboratory, Department of Physics, University of California at Berkeley, Berkeley, California 94720, USA}

\author{Irwin Huang}
\affiliation{Department of Physics and Astronomy, University of Rochester, Rochester, NY, USA}

\author{Bibek Bhandari}
\affiliation{Institute for Quantum Studies, Chapman University, Orange, CA, USA}

\author{Nicholas E. Frattini}
\email{Present address: Nord Quantique, Sherbrooke, QC J1J 2E2, Canada.}
\affiliation{Department of Applied Physics and Physics, Yale University, New Haven, CT 06520, USA}

\author{Shruti Puri}
\affiliation{Department of Applied Physics and Physics, Yale University, New Haven, CT 06520, USA}

\author{Justin Dressel}
\affiliation{Institute for Quantum Studies, Chapman University, Orange, CA, USA}

\author{Andrew N. Jordan}
\affiliation{Department of Physics and Astronomy, University of Rochester, Rochester, NY, USA}
\affiliation{Institute for Quantum Studies, Chapman University, Orange, CA, USA}
\affiliation{The Kennedy Chair in Physics, Chapman University, Orange, CA 92866, USA}

\author{David I. Santiago}
\affiliation{Computational Research Division, Lawrence Berkeley National Laboratory, Berkeley, California 94720, USA}

\author{Irfan Siddiqi}
\affiliation{Quantum Nanoelectronics Laboratory, Department of Physics, University of California at Berkeley, Berkeley, California 94720, USA}
\affiliation{Computational Research Division, Lawrence Berkeley National Laboratory, Berkeley, California 94720, USA}

\date{\today}

\begin{abstract}

The Kerr-cat qubit is a bosonic qubit in which multi-photon Schr$\ddot{\text{o}}$dinger cat states are stabilized by applying a two-photon drive to an oscillator with a Kerr nonlinearity. The suppressed bit-flip rate with increasing cat size makes this qubit a promising candidate to implement quantum error correction codes tailored for noise-biased qubits. However, achieving strong light-matter interactions necessary for stabilizing and controlling this qubit has traditionally required strong microwave drives that heat the qubit and degrade its performance. In contrast, increasing the coupling to the drive port removes the need for strong drives at the expense of large Purcell decay. By integrating an effective band-block filter on-chip, we overcome this trade-off and realize a Kerr-cat qubit in a scalable 2D superconducting circuit with high coherence. This filter provides 30 dB of isolation at the qubit frequency with negligible attenuation at the frequencies required for stabilization and readout. 
 We experimentally demonstrate quantum non-demolition readout fidelity of 99.6$\%$ for a cat with 8 photons. Also, to have high-fidelity universal control over this qubit, we combine fast Rabi oscillations with a new demonstration of the $X(\pi/2)$ gate through phase modulation of the stabilization drive. Finally, the lifetime in this architecture is examined as a function of the cat size of up to 10 photons in the oscillator achieving a bit-flip time higher than 1 ms and only a linear increase in the phase-flip rate, in good agreement with the theoretical analysis of the circuit. Our qubit shows promise as a building block for fault-tolerant quantum processors with a small footprint.

\end{abstract}
\maketitle 
\section{Introduction}
  \begin{figure*}[tp]

    \includegraphics[width=17.5 cm]
{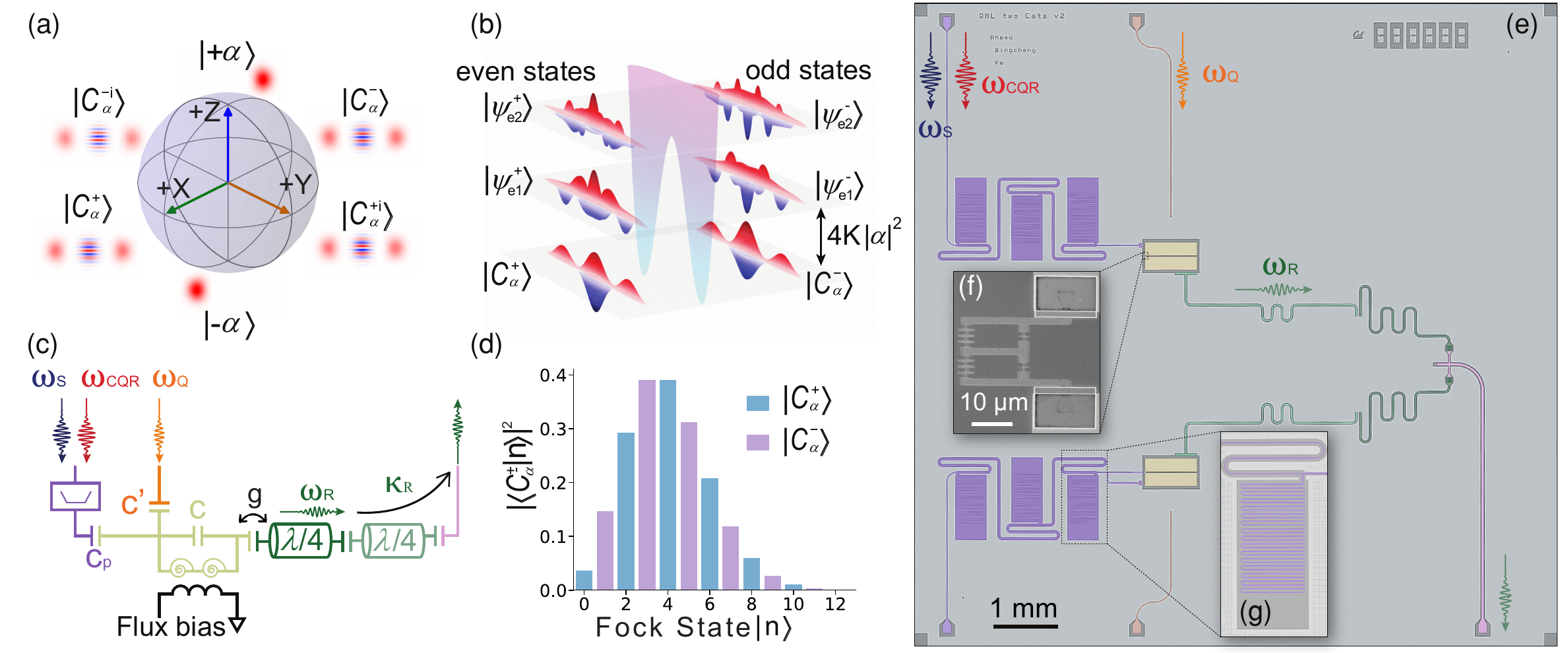}
    \caption{ Kerr-cat qubit encoding and 2D chip architecture. (a) The Bloch sphere for the Kerr-cat qubit is spanned by the cat states along the x-axis and the coherent states along the z-axis. (b) The energy pseudopotential is a double well potential at $\pm \alpha$ with energy gap $4K\alpha^2$ separating the computational space from the excited states. The Wigner function of the cat states and the excited states are shown according to the parity. (c) The circuit diagram shows the nonlinear oscillator with the ports used for stabilization ($c_p$), control ($c'$), and readout. The numerical values of the design parameters are reported in Table \ref{table:1}. (d) The distribution of the cat states with an average photon number equal to 4 (i.e., $|\alpha|^2=\langle  \hat{a}^{\dagger}\hat{a}\rangle	$) in terms of the Fock states showing the parity of the cat states and the tail of the distribution. (e) The 2D chip (10x10 mm$^2$) includes two uncoupled SNAILmons (yellow), each with its dedicated readout resonator and Purcell filter (light and dark green). Resonant drives at frequency $ \omega_\text{Q}$ are delivered through weakly coupled control lines (orange), while the strong off-resonant drives used for stabilization ($\omega_\text{S}$) and cat readout ($\omega_\text{{CQR}}$) are delivered through a strongly coupled port with a band-block filter (purple). (f) The nonlinearity of the oscillator comes from two SNAIL loops biased around $\Phi_\mathrm{ext}/\Phi_\mathrm{0} = 0.3$ by an off-chip flux coil. (g)  Each stub of the filter consists of a meander section to have destructive interference around the qubit frequency.
}
    \label{fig1}
\end{figure*}
Superconducting quantum circuits offer a rich platform for engineering Hamiltonians that encode and manipulate quantum information faster than the decoherence rates \cite{Transmon_intro, Transmon_intro_theory, Fluxonium_intro, Fluxonium_intro_coherence, ZeroPi, Zero_pi_theory, GKP_theory, GKP_QEC,siddiqi2021engineering, CurrentState_SupercondcutingQubits}. Integrating the Josephson junction into those circuits enables a wide range of nonlinear terms in the Hamiltonian that can be used to implement multi-photon wave mixing processes in the microwave range \cite{Rev_CQED_Blais, QuantumEngineerGuide}. 

In particular, the family of cat qubits uses this nonlinearity to stabilize the superposition of macroscopically distinct so-called Schrödinger cat states. Depending on the details of the encoding scheme, those qubits can exhibit either exponential protection or parity changes with single-photon loss, making them a valuable resource for implementing quantum error correction schemes \cite{vlastakis2013deterministically, gao2019entanglement, ofek2016extending,wang2016schrodinger, song2019generation, PhysRevX.13.021004, Dissipative_cat_intro,First_cat, QEC_squeezed_cat_qubits,traversing_the_bosonic_ladder,Atomic_cat,Atomic_cat2}. One approach to stabilize those states with autonomous protection from bit-flips is by engineering two-photon addition and subtraction in the case of the dissipative cat. This approach offers exponential protection with increasing cat size but requires external ancillas to engineer the dissipation and perform readout \cite{Dissipative_cat_intro, Dissipative_cat_intro2, Dissipative_cat_intro3, Dissipative_cat_intro4}. 

An alternative approach to stabilize the cat states is by subjecting a Kerr nonlinear oscillator to a coherent two-photon drive \cite{Shruti_KerrCat_intro, Goto_KerrCat_intro}. This approach has the advantage of creating a relatively large energy gap (5-100 MHz) between the computational space and the rest of the Hilbert space, enabling fast logic operations (20-$400$ ns) and high-fidelity readout with no added ancillas. The Kerr-cat qubit gained a lot of attention in terms of examining  its noise-biased behavior both theoretically \cite{Shruti_bias_preserving_gates, Shruti_error_syndrom, Theory_detuning_lifetime, critical_cat, Cats_Confinment} and experimentally \cite{First_cat,frattini2022squeezed, Fastgeneration, SQUID_KerrCat, Jaya_detuning_experiment}. Also, this qubit makes a promising candidate to implement the XZZX surface code provided it can be realized in a scalable architecture \cite{XZZX_paper, XZZX_paper2, XZZX_original}. 

The protection against single-photon loss in this qubit can be intuitively understood by noting that the coherent states $\ket{\pm\alpha}$ are quasi-orthogonal eigenstates of the annihilation operator ($\hat{a}$) with eigenvalues $\pm\alpha$, so the dominant process of single-photon loss only leads to a linear increase in the qubit dephasing at the benefit of exponential suppression in the relaxation rate. The first realizations of this qubit were in 3D architectures and relied  on strong microwave drives which led to notable heating and degraded the qubit lifetime \cite{First_cat,frattini2022squeezed, Jaya_detuning_experiment}. Therefore, it is still an open question to see whether possible to realize large cat states with high coherence in 2D chips while maintaining high-fidelity operations. This uncertainty arises because the 3D environment is known to result in longer lifetimes but weaker light-matter coupling due to the larger spread of the  electromagnetic field in the extra dimension \cite{3D_transmon, Blais_JC, wallraff2004strong}.  

Realizing a 2D circuit with strong light-matter coupling without sacrificing the lifetime is crucial to scaling up the Kerr-cat qubit.  The inherent trade-off is that coupling the drive port strongly to the qubit (i.e., large coupling capacitance)  leads to large Purcell decay. In contrast, coupling weakly at the expense of strong driving heats the qubit and degrades its performance, as shown in other strongly driven qubits \cite{Long_heating, Dissipative_cat_intro2}.

In this work, we overcome this challenge by using an effective band-block filter that enables strong coupling off-resonance (at the frequency of the stabilization drive $\omega_\text{S} $ and the cat  readout drive $\omega_{\text{CQR}}$) while suppressing the qubit Purcell decay rate orders of magnitude smaller than the limit imposed by materials defects. Operating at a Kerr nonlinearity of 1.2 MHz, we observe improvements in the bit-flip time with increasing  drive strength up to cat of size 10 photons. The bit-flip time exceeds 1 ms while only increasing the phase-flip rate linearly with the cat size. Also, the large energy gap at this Kerr value enabled us to perform the highest fidelity readout and gates on Kerr-cats to date. Our work, therefore, provides a path for scaling noise-protected qubits toward fault-tolerant quantum computers.
\section{Encoding}
The Kerr-cat qubit has the following Hamiltonian in the rotating frame \cite{Shruti_KerrCat_intro, Goto_KerrCat_intro}:
\begin{equation}\label{eq1_main}
\hat{H}_{\text{KC}}/\hbar=-K \hat{a}^{\dagger2}\hat{a}^2+\epsilon_2\hat{a}^{\dagger2 }+\epsilon^*_2\hat{a}^2 
\end{equation}
  where $K$ is the strength of the Kerr nonlinearity, $\epsilon_2$ is the amplitude of the two-photon drive (i.e., the stabilization drive), and $\hbar$ is the reduced Planck constant. This Hamiltonian stabilizes two coherent states  $\ket{\pm\alpha}$ provided the cat size $\alpha^2=\epsilon_2/K$. The cat states are the orthogonal even/odd parity states formed by the superposition of the coherent states $\ket{C_\alpha^{\pm}}=\frac{1}{\sqrt{1\pm e^{-2\alpha^2}}}\frac{1}{\sqrt{2}}(\ket{\alpha}\pm \ket{-\alpha} ) $ and span the Bloch sphere along the x-axis as shown in Fig. \ref{fig1}(a). The orthogonal states along the z-axis are approximately the coherent states (i.e., $\ket{\pm Z}=\frac{1}{\sqrt{2}}(\ket{C_\alpha^+}\pm\ket{C_\alpha^-})\approx \ket{\pm \alpha}$). The energy pseudopotential of this Hamiltonian is a double-well with an energy gap $E_\text{{gap}}/\hbar\approx4K\alpha^2$ \cite{Shruti_bias_preserving_gates} separating the computational space from the rest of the Hilbert space as shown in Fig. \ref{fig1}(b). 
  
This Hamiltonian is now traditionally realized by charge pumping a SNAIL (Superconducting Nonlinear Asymmetric Inductive eLement) oscillator with a coherent drive at approximately twice the oscillator (i.e. qubit) frequency $\omega_\text{S}\approx 2 \omega_\text{Q}$ \cite{First_cat,frattini2022squeezed}. Utilizing the third-order nonlinearity of the SNAIL $\hat{\varphi}^3$ (where $\hat{\varphi}$  is the superconducting phase operator), a photon from the drive is consumed to create two photons in the oscillator (${a}^{\dagger2 }$) at the rate $\epsilon_2$; this is accompanied by the complementary process of eliminating two photons from the oscillator ($\hat{a}^2$) and creating a pump photon \cite{SNAIL_intro_Nick, Nick_SNAIL_Optimization, Floquet_SNAIL_KerrCat}. The other nonlinearity of the SNAIL is a fourth-order ($\hat{\varphi}^4$), which in the rotating wave approximation (RWA) adds the Kerr nonlinearity $-K\hat{a}^{\dagger2}\hat{a}^2$ to the Hamiltonian. We refer to this element in the Fock encoding as the SNAILmon (i.e., transmon with small anharmonicity provided by the SNAIL).
\section{Design}
The 2D  superconducting circuit used to realize the Kerr-cat qubit with the SNAILmon is shown in  Fig. \ref{fig1}  (e-g). The design of the SNAILmon contains two large niobium capacitor paddles and is based on high-coherence transmon qubit designs \cite{Berkeley_transmon, kreikebaum2020improving}. In addition, our design contains two SNAILs consisting of aluminum Josephson junctions fabricated with a bridge-free two-angle deposition process. Concatenating multiple SNAILs has allowed large cat sizes in previous Kerr-cat qubit experiments \cite{frattini2022squeezed}. The readout circuitry contains a readout resonator coupled to a Purcell filter  (green), which allows for fast readout while protecting the qubit from Purcell decay. For control, a weakly coupled port ($c'$ in Fig. \ref{fig1}(c)) is used for on-resonance driving. For stabilization and readout, a strongly coupled port is used for efficient pumping ($c_p$ in Fig. \ref{fig1}(c)) with an effective band-block filter to suppress the Purcell decay. 

The nonlinearities of our circuit can be tuned with an external flux piercing each SNAIL. In this work, we fix the external flux to realize a Kerr non-linearity of $K/2\pi=1.2$ MHz, a value between the first realization of the Kerr-cat (6.7 MHz) \cite{First_cat} and the smaller value (0.3 MHz) in \cite{frattini2022squeezed, Jaya_detuning_experiment}. At this operating point, we observe an improvement in the bit-flip time with increasing cat size while maintaining the ability to  perform fast logic gates and coherent mapping between the Fock qubit and the cat qubit if needed (see Fig. \ref{Fig1_supp} in Appendix \ref{mapping}). 
\begin{figure}[tp]
    \centering
    \includegraphics[width=8cm]{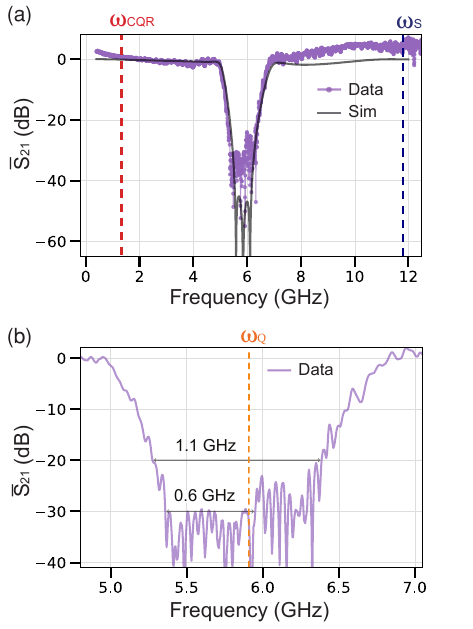}
    \caption{The transmission parameter of the wide band-block filter used for strong qubit-drive coupling. (a) The measured performance for the full range of the filter  (purple). The dark grey line is the simulated performance using finite-element simulation. The filter has minimum attenuation at the frequency of the stabilization drive $\omega_\text{S}$ and the cat quadrature readout $\omega_{\text{CQR}}$. At the qubit frequency $\omega_\text{Q}$, the filter provides very high isolation of 30 dB. (b) Shows a zoomed view around the qubit frequency with the 20,30 dB attenuation ranges, which makes this filter suitable for multi-qubit chips with strong coupling between the qubits and large detuning due to its large bandwidth.}
    \label{Fig4}
\end{figure}
\begin{figure*}[htp]

    \includegraphics[width=17.5cm]
{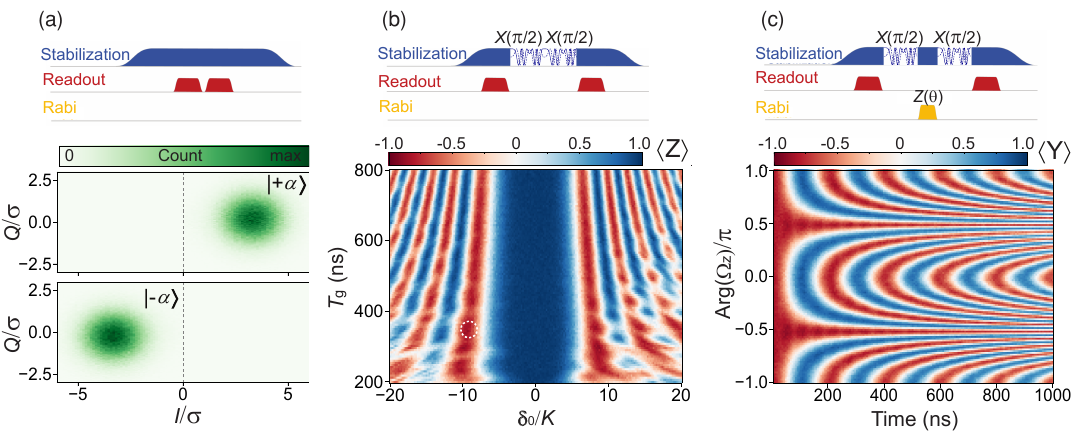}
    \caption{Readout and gates with the Kerr-cat qubit. (a) The sequence to measure the QNDness of the cat quadrature readout involves two consecutive measurements to determine the probability of getting the same outcome. The demodulated output signal for the second measurement is shown in the IQ plane conditional on the first measurement being $\ket{\pm \alpha}$. The data is normalized with respect to the standard deviation of the Gaussian distribution ($\sigma$) of each output. For every initial state we average $10^7$ measurements. (b) The calibration of the detuning $X(\pi/2)$ gate is based on using two $X(\pi/2)$ pulses to flip the qubit and change the sign of $\left<Z\right>$. This gate is implemented by deforming the double-well potential, allowing the two coherent states to tunnel. The presence of the derivative in Eq.\ref{eq_detuning}  enables one to detune the phase in either direction. The calibration consists of finding the best choice of gate time $T_g$ and maximum modulation depth $\delta_0$ to maximize the population transfer between the coherent states. The dashed circle indicates the calibration point. (c) The calibration of the drive phase to implement the Z gate is determined by observing the fastest oscillations in $\left<Y\right>$ (plotted relative to the phase of $\alpha$). When driving in phase, we get an enhancement factor $2\alpha$ while driving out of phase leads to exponentially suppressed rates.}
    \label{fig3}
\end{figure*}

The band-block filter we use to couple strongly to the stabilization and readout drives is shown in Fig. \ref{fig1}(e, g). This filter consists of multiple stubs, each leading to destructive interference of the microwave signal around the qubit frequency. This is achieved by adjusting the length of the meandering section (i.e., the quarter wave resonators) and simulating the scattering parameters in Ansys-HFSS \cite{Pozar}. Ideally, this filter has high isolation at the qubit frequency ($\omega_\text{Q}/2\pi=5.9$ GHz) with zero attenuation at the frequency of the stabilization drive ($\omega_\text{S}/2\pi=11.8$ GHz) and the frequency of the cat quadrature readout ($\omega_{\text{CQR}}/2\pi=1.2$ GHz). In Fig. \ref{Fig4}, we present an independent measurement of the transmission coefficient $\Bar{S}_{21}$ of the filter in a dedicated chip (for more details about the setup see Appendix \ref{filter}). In Fig. \ref{Fig4}(a), we show the behavior of the normalized scattering parameter $\bar{S}_{21}$ for the full range of interest. The 30 dB of isolation at the qubit frequency  corresponds to a three-order of magnitude improvement in the lifetime, pushing the Purcell limit through the strongly coupled port to 12 ms, which is much higher than the measured single photon decay time of the Fock qubit of  $T_1= 38.5$ $\mu\text{s}$. Material defects in the interfaces have been shown to limit the lifetime of 2D architectures at this range. \cite{Surface_treatments_1, Surface_treatments_2, Surface_treatments_3,Surface_treatments_4, Surface_treatments_5,Nb_cavity_TLS}. 

In Fig. \ref{fig3}(b), we zoom in around the qubit frequency to examine the range of 20 dB and 30 dB of isolation. The large bandwidth of this filter makes it suitable for multi-qubit chips with strong qubit-qubit coupling while maintaining a negligible Purcell effect and efficient pumping for all the qubits. This filter can also be incorporated in any circuit QED experiment in which a strong off-resonance drive is required to implement the desired interaction \cite{SNAIL_Coupler, SNAIL_Coupler2, SNAIL_Coupler3}. The idea of integrating on-chip filters to enable strong coupling with mitigated Purcell effects was first introduced in the context of fast readout \cite{First_Purcell, Purcell_readout, QuantumEngineerGuide}, and the use of band block filters to enable strong coupling off-resonance was also explored more recently \cite{Bandblock_filter_subharmonic_driving, Bandblock_filter_subharmonic_driving_2}. 

\section{Qubit Control and Readout}\label{Qubit Control and Readout}
To create the Kerr-cat qubit, we ramp up the stabilization drive to transform the SNAILmon Hamiltonian to the Kerr-cat qubit Hamiltonian in the rotating frame (see Appendix \ref{making_cats_from_snails} for derivation). In this section, we focus on characterizing the readout and the single qubit gates for a cat of size $\alpha^2=$4 (see Appendix \ref{calibration_supp} for more details). Starting with the cat quadrature readout in Fig. \ref{fig3}(a), we can measure the qubit along the z-axis by sending a microwave pulse to the qubit at frequency $\omega_{\text{CQR}}/2\pi=(\omega_\text{R}-\omega_\text{S}/2)/2\pi= 1.2$ GHz which implements a beam-splitter interaction between the qubit and readout mode $\hat{H}_{\text{CQR}}\approx\epsilon_{\text{CQR}} \alpha \hat{Z}(\hat{b}+\hat{b}^{\dagger})$ which is derived in Appendix \ref{CQR_sub}  \cite{First_cat}. The output signal is a coherent state with one of two opposite phases according to the state of the qubit, as shown in Fig. \ref{fig3}(a). For a cat of size 4 and $4$ $ \mu$s readout time, we measure a QNDness  $\mathcal{Q}=(p(+\alpha|+\alpha )+p(-\alpha|-\alpha ))/2$ of $98.3 \%$ where $p(+\alpha|+\alpha )$ is the probability that the second measurement gives $\ket{+\alpha}$ consistent with the first measurement result. This high QND readout enables us to use the measurement to initialize the qubit in the Z basis. For a cat of size 8, the QNDness of the readout increases to 99.6\%, as shown in Fig. \ref{Fig1_supp}. Those values are equal to the state-of-the-art across all superconducting qubits \cite{High_QND_RO_1,frattini2022squeezed, High_QND_RO_2}. 
    \begin{figure*}[tp]
    \includegraphics[width=17.5 cm]
    {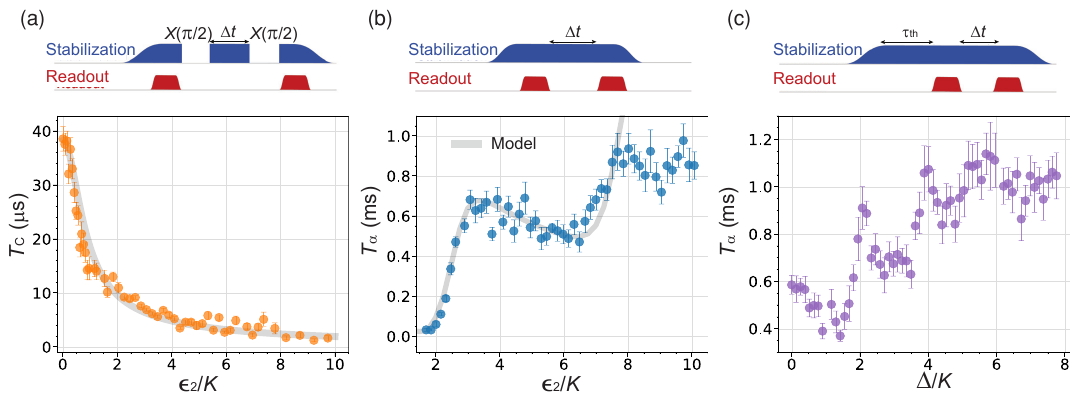}
    \caption{The lifetime of the Kerr-cat qubit. (a) The sequence to measure the lifetime of the cat states requires ramping up the stabilization drive in $\tau_{\text{ramp}}=3\mu$s to keep the population in the cat space. The first readout is used to initialize the qubit in the Z-basis. Two $X(\pi/2)$ gates are used to prepare and measure the lifetime of the cat states $T_{\rm C}$ (i.e., the phase-flip  time) for every cat size $\alpha^2=\epsilon_2/K$. The lifetime decreases with the cat size,  starting with a lifetime of around $38.5 \mu s$, consistent with the single-photon loss being the dominant factor. (b) The lifetime of the coherent states $T_\alpha$ (i.e., bit-flip time) follows a staircase-like behavior with peaks around cat size $\alpha^2=3, 8$ and a lifetime of  650$\mu$s and 950$\mu$s, respectively. The data is fitted to a master equation model, which includes single and multi-photon heating effects and frequency fluctuations, as explained in the main text and Appendix \ref{Master_equation}. (c) The sequence to prepare the detuned cat includes extra wait time $\tau_{\rm th}= 35$ $\mu$s to put the qubit in a thermal state and use the first readout to prepare the qubit. The data shows improvement in the lifetime of the coherent states for a stabilization drive of strength  $\epsilon_2/K =4$  as we introduce a red detuning to the drive in units of the Kerr nonlinearity. The value peaks around even multiples of $K$ due to the acquired degeneracy of the excited states.}
    \label{fig2}
\end{figure*}

Universal single-qubit control typically requires driving on resonance to do rotations along two orthogonal axes, depending on the phase of the drive \cite{QuantumEngineerGuide, Shruti_KerrCat_intro}. However, this protocol in the cat encoding leads to an enhancement factor $2\alpha$ in the speed of the $Z(\theta)$ rotations and an exponential suppression of the $Y(\theta)$ rotations \cite{First_cat}(For more details, check Appendix \ref{calibration_supp}). To overcome this challenge, it is sufficient to use the $Z(\theta)$ rotations with $X(\pi/2)$ gates to have universal control. The first proposal to do an $X(\pi/2)$ gate was demonstrated in \cite{First_cat} by abruptly turning off the stabilization drive for a period $\pi/2K=216$ ns. However, this protocol removes the protection the stabilization drive adds to the qubit, and it tends to have worse performance for large cats \cite{frattini2022squeezed}. 

To improve on this implementation, we demonstrate a new $X(\pi/2)$ gate by phase modulating the stabilization drive according to  $\epsilon_2 \rightarrow \epsilon_2 e^{-it\delta_d(t)}$. This modulation modifies the double well potential and allows the coherent states to tunnel. With the appropriate unitary transformation, the phase modulation of the drive adds the following term to the Kerr-cat Hamiltonian:
\begin{equation}
H_d(t)=-\frac{1}{2}(\delta_d(t)+t\dot{\delta}_d(t)) \hat{a}^{\dagger}\hat{a}
\label{eq_detuning}
\end{equation}
Fig. \ref{fig3}(b) shows the expectation value of Pauli Z after implementing two $X(\pi/2)$ pulses as a function of the maximum detuning $\delta_0$ in the phase modulation waveform and the gate time $T_g$. This gate implementation is resource-efficient since it does not require additional tones and does not disturb the SNAIL parameters compared to a fast DC flux. The coherent tunneling between the states inside the two energy wells using a controlled time-dependent detuning to perform logic gates was first examined theoretically  \cite{Goto_KerrCat_intro, Holonomic_gate, Shruti_KerrCat_intro}, and the static detuning was shown to improve the lifetime 
experimentally \cite{Jaya_detuning_experiment, frattini2022squeezed}. The simulation of this gate according to Eq. \ref{eq_detuning} is shown in Fig. \ref{Fig_xgate} (more details in Appendix \ref{X gate supp}). From the chevron plot, we chose $T_g=320$ ns and a detuning  $\delta_0/K=-8.2$ for maximum population transfer. 

Next, we calibrate the phase of the drive to implement $Z(\theta)$ rotations by driving on resonance. The added term to the Kerr-cat Hamiltonian has the form $\approx 2\alpha |\Omega_{\rm z}|\cos(\theta_{\rm z})\hat{Z}-2\alpha \Omega_{\rm z} \sin(\theta_{\rm z}) e^{-2|\alpha|^2} \hat{Y}$ as long as the drive amplitude $\Omega_{\rm z}$ is small compared to the energy gap $E_{\text{gap}}/\hbar$ \cite{Shruti_bias_preserving_gates, First_cat}. Fig. \ref{fig3}(e) shows the protocol to calibrate the phase of the drive ($\text{Arg}(\Omega_{\rm z })=\theta_{\rm z}$) by leveraging the cat quadrature readout and the $X(\pi/2)$ gate introduced earlier. The second $X(\pi/2)$ gate enables us to measure Rabi oscillations in $\left<Y\right>$ as we sweep the phase of the Rabi drive $\theta_{\rm z}$ with maximum contrast $|\left<Y\right> |\approx 0.9$. When using the traditional $X(\pi/2)$ gate where the stabilization drive is turned off \cite{First_cat}, the maximum contrast drops to $|\left<Y\right> |\approx 0.5$. We can see in Fig. \ref{fig3}(c) that driving in phase leads to the maximum Rabi rate while driving around $\theta_{\rm z}= \pm \pi/2$ leads to suppressed rates.
For a cat of size 4, we choose a gate time of $120$ ns to implement the $Z(\pi/2)$ gate. With the $X(\pi/2)$ and the $Z(\theta)$ calibrated, we have a universal set of single-qubit gates. Using process tomography, we measure gates fidelities  $\mathcal{F}_{X_{\pi/2}}=90.8\%$ and $\mathcal{F}_{Z_{\pi/2}}=91.7\%$. For the identity, we measure $\mathcal{F}_{I}=92.7\%$, which implies that state preparation errors are the limiting factor. The Pauli transfer matrices for those processes with the details of the analysis are provided in Appendix \ref{tomog}.
\section{Cat Qubit Lifetime}
We start by measuring the lifetime of the cat qubit as a function of the cat size $\alpha^2=\epsilon_2/K$ by varying the strength of the two-photon drive $\epsilon_2$ as shown in Fig. \ref{fig2}. Every sequence begins with a readout signal to initialize the qubit in the Z-basis. To measure the lifetime of the cat states $T_{\rm C}$ (i.e., phase-flip time), we use the Kerr $X(\pi/2)$ gate introduced in \cite{First_cat}, since it does not depend on the cat size in the calibration. For every cat size, we vary the delay time $\Delta t$ to observe the exponential decay of the coherence and perform another $X(\pi/2)$ since the cat readout projects the qubit to the Z-basis. The measured lifetime in Fig. \ref{fig2}(a)  follows the trend in grey, which is the linear trade-off for large cats described by $T_{\rm C}=T_1 / 2\langle \bar{n} \rangle   $
with $\langle \bar{n} \rangle =|\alpha|^2 (1+e^{-4|\alpha|^2})/(1-e^{-4|\alpha|^2})$  the time-averaged number of photons in the qubit \cite{frattini2022squeezed}. 

The linear increase in the phase-flip rate of the cat states should come at the benefit of exponential protection against the single photon loss in the bit-flip rate. In Fig. \ref{fig2}(b), we examine the lifetime of the coherent states $T_\alpha$ (i.e. bit-flip time) as a function of the cat size. The staircase shape we observe implies that thermal excitations limit the lifetime. A bit-flip error can happen through tunneling to the higher energy states, which starts to become asymptotically degenerate at the points where the lifetime increases sharply with the cat size. This trend was first observed in \cite{frattini2022squeezed}. The staircase behavior is captured by heating effects in the master equation like $\kappa_{\omega_d/2} n_{\omega_d/2}\mathcal{D}\{\hat{a}^\dagger\}$ while the exact location of the plateaus in this model depends on the spectrum of the dephasing noise $\kappa_\phi(\omega)$ \cite{Jaya_SNAIL_theory} (see Appendix \ref{Master_equation} for more details). Our data up to cats of size 8 can be understood in terms of the single and multi-photon heating and dephasing of the SNAIL with thermal population $n_{\omega_d/2}=2\%$. This value is the lowest compared to other Kerr-cat experiments or similar strongly driven qubits\cite{Long_heating}. However, this is still higher than the upper bound of $0.3\%$   value we measure from the residual population of the $\ket{n=1}$ state in the Fock basis \cite{Qubit_temp_MIT}. The details of the master equation model are presented in Appendix \ref{Master_equation}. 

Next, we focus on cats of size 4, which offer the most gain in $T_\alpha$ without sacrificing $T_{\rm C}$ much. It is also possible to red-tune the stabilization drive to boost the lifetime of the coherent states by adding the harmonic term $\Delta \hat{a}^{\dagger}\hat{a}$ with $\Delta=\omega_\text{Q}-\omega_\text{S}/2$ to the Kerr-cat Hamiltonian in Eq.\ref{eq1_main} \cite{frattini2022squeezed, Jaya_detuning_experiment}. Fig. \ref{fig2}(c) shows the improvement in $T_\alpha$ for $\epsilon_2/K=4$  as we sweep the detuning. The lifetime peaks around even multiples of $K$ due to the Hamiltonian picking extra degeneracies in the excited states \cite{Jaya_detuning_experiment}. However, in the case of the detuned cat, the ground state of the Fock qubit $\ket{n=0}$ does not map to the even cat state of the detuned cat qubit, so instead, we keep the stabilization drive for a time $\tau_{\text{th}}\approx T_1$
necessary to put the detuned cat in a thermal state, then we use the cat quadrature readout to initialize the qubit.  Those results demonstrate the potential of the Kerr-cat qubit to benefit from the suppressed bit-flip rate with only a linear increase in the phase-flip rate with increasing cat size while maintaining high-fidelity operations.
\section{Conclusion }
Our results demonstrate a high-coherence Kerr-cat qubit in a scalable 2D architecture with high-fidelity single-qubit operations and minimal heating. This demonstration is enabled by integrating a wide band-block filter for efficient energy delivery, enabling strong light-matter coupling with negligible Purcell effect. We report improvement in the bit-flip time up to cats of size  $\alpha^2=10$  with lifetimes higher than 1 ms, in good agreement with our theoretical modeling of the circuit. Our qubit was based on a SNAILmon oscillator with a single photon decay time $T_1=38.5 $ $\mu s$.  

Operating at a Kerr nonlinearity of 1.2 MHz, we get an energy gap $4K \alpha^2=$  9.6-48 MHz for cats of size 2-10, which enables fast operation compared to the decay rates. For those device parameters, we demonstrate quantum non-demolition readout with QNDness of 98.3\% for a cat of size 4 and 99.6\% for a cat of size 8. We also introduce a resource-efficient way to implement the $X(\pi/2)$ gate with phase modulation of the stabilization drive. Combined with continuous and fast $Z(\theta)$ rotations, this is sufficient to have universal control over the Kerr-cat qubit. For a cat of size 4, we implement single qubit $Z(\pi/2)$ in 120 ns and the $X(\pi/2)$ in 320 ns with gate fidelities of $91.7\%$ and $90.8\%$ respectively, limited by state preparation using process tomography. A detailed analysis of those gates in terms of the noise bias without SPAM errors is the subject of future work. 

With the goal of implementing quantum error correction codes with many physical Kerr-cat qubits the problem of finding an efficient pumping scheme that does not heat the cryogenic environment is of most importance. Our solution relied on integrating an effective band-block filter to remove the trade-off between strong capacitive coupling to the qubit and large Purcell decay to enable strong light-matter interactions. With 30 dB of isolation at the qubit frequency, large bandwidth, and negligible attenuation at the pump frequencies, this design enables the realization of multi-qubit processors.

In conclusion, we introduced a novel scheme for efficient power delivery in 2D architecture to realize the Kerr-cat qubit with high coherence, fast logic gates, and state of art QND readout.
This work extends the reach of the Kerr-cat qubit to strongly coupled multi-qubit 2D processors, which can be used to examine the fidelity of the noise-bias-preserving two-qubit CZ and CX gates \cite{Shruti_bias_preserving_gates} and quantum error correction schemes tailored for noise-biased qubits  with Kerr cats \cite{XZZX_original, XZZX_paper, XZZX_paper2}. Future single-qubit experiments can focus on increasing the Kerr nonlinearity of the qubit without jeopardizing the lifetime to enable faster single-qubit gates with improved fidelity and examine the root causes of the saturation in the lifetime with controlled dissipation.  

\section{Acknowledgments}
 We thank Chuan-Hong Liu, Akel Hashim, and Ravi Naik for valuable discussions on the experiment. We are grateful to Alexei Marchenkov and Alexander Tselikov for the blueprint of the on-chip RF filter. This work was supported by the U. S. Army Research Office under grant W911NF-22-1-0258.

\input{supplement}

\bibliography{thebibliography}

\end{document}

%% file: supplement.tex
\title{Supplementary Information }
\appendix
\section{Realizing the Kerr-cat Hamiltonian From the SNAIL} \label{making_cats_from_snails}
\subsection{The Kerr-cat Hamiltonian}
The Kerr-cat Hamiltonian $\hat{H}_{\text{KC}}$ can be realized by adding a stabilization drive $\hat{H}_{S}$ at twice the SNAILmon frequency to the SNAIL Hamiltonian ($\hat{H}_{a}$) and moving to a rotating frame as follows ($\hbar$=1) \cite{First_cat}:
\begin{equation}
\hat{H}_0=\hat{H}_{a}+\hat{H}_{S},
\end{equation}
\begin{equation}
\hat{H}_{a}=\omega_{a,0}\hat{a}^{\dagger}\hat{a}+g_3 (\hat{a}^{\dagger}+\hat{a})^3+g_4 (\hat{a}^{\dagger}+\hat{a})^4,
\end{equation}
\begin{equation}
\hat{H}_{S}=2 \text{Re}(e^{i\omega_{\text{s}}t})(\epsilon_{\rm s,0} \hat{a}^{\dagger}+\epsilon^*_{\rm s,0}\hat{a}).
\end{equation}
The operator $\hat{a} $ is the bosonic annihilation operator of excitations in the SNAILmon oscillator. The oscillator frequency is $\omega_{\rm a,0}$, and third, and fourth-order nonlinearities are $g_3$  and $g_4$, respectively. Now, to eliminate the time dependence, we apply two consecutive transformations. First, we move to a displaced Fock basis characterized by $-\xi_s$ and then to a rotating frame characterized by $\omega_{\rm r}$
\begin{equation}
\hat{U}= \hat{U}_{\text{rot}}  \hat{U}_{\text{dis}}=e^{i\omega_\text{r}t\hat{a}^{\dagger}\hat{a}} \cdot e^{\xi_\text{s} \hat{a}^{\dagger}-\xi^*_\text{s} \hat{a}}
\end{equation}
This leads to the effective Hamiltonian $\hat{H}_{\text{KC}}=\hat{U}\hat{H}_0\hat{U}^\dagger+i\dot{\hat{U}}\hat{U}^\dagger$. The second term in $\hat{H}_{\text{KC}}$ eliminates the harmonic part in $\hat{U}\hat{H}_0\hat{U}^\dagger$ if $\omega_\text{r}=\omega_{a,0}$. More precisely: 
 \begin{equation}
i\dot{\hat{U}}\hat{U}^\dagger=-\omega_{\text{r}} \hat{a}^{\dagger}\hat{a}+ie^{i\omega_{\text{r}}t \hat{a}^{\dagger}\hat{a}}(\dot{\xi}_{\text{s}} \hat{a}^{\dagger}-\dot{\xi}^*_{\text{s}} \hat{a})e^{-i\omega_{\text{r}}t \hat{a}^{\dagger}\hat{a}}
 \end{equation}
 
The impact of sandwiching $\hat{H}_0$ is summarized by first displacing $a$ ($a^{\dagger} $) by $\xi_{\text{s}}$ ( $\xi_{\text{s}}^*$)  and then adding a time-dependent  factor $e^{-i \omega_{\text{r}}t} (e^{i \omega_{\text{r}}t}) $, which is crucial when we apply the rotating wave approximation (RWA) as follows:
\begin{equation}
\hat{U}_{\text{dis}}\ \hat  {a} \ \hat{U}^\dagger_{\text{dis}}=\hat{a}-\xi_{\text{s}}
\end{equation}
\begin{equation}
\hat{U}_{\text{dis}}\ \hat  {a}^\dagger\ \hat{U}^\dagger_{\text{dis}}=\hat{a}^\dagger-\xi^*_{\text{s}}
\end{equation}
\begin{equation}
\hat{U}_{\text{rot}}\ \hat  {a} \ \hat{U}^\dagger_{\text{rot}}=\hat{a} e^{-i \omega_{\text{r}}t}
\end{equation}
\begin{equation}
\hat{U}_{\text{rot}}\ \hat  {a}^\dagger \ \hat{U}^\dagger_{\text{rot}}=\hat{a}^\dagger e^{i \omega_{\text{r}}t}
\end{equation}
Now we can expand $\hat{H}_{\text{KC}}$ to get:
\begin{equation}
\begin{split}
    \hat{H}_{\text{KC}}&=-\omega_{\text{r}} \hat{a}^{\dagger}\hat{a} +i\dot{\xi}_{\text{s}}\hat{a}^\dagger e^{i \omega_{\text{r}}t}-i\dot{\xi}^*_{\text{s}} \hat{a} e^{-i \omega_{\text{r}}t}\\ &+\omega_{a,0} (\hat{a}^\dagger e^{i \omega_{\text{r}}t}-\xi^*_s) (\hat{a} e^{-i \omega_{\text{r}}t}-\xi_{\text{s}}) \\
    &+2 \text{Re}(e^{i\omega_{\text{s}}t})[\epsilon_{\rm s,0} (\hat{a}^\dagger e^{i \omega_{r}t}-\xi^*_\text{s})+\epsilon^*_{\rm s,0}(\hat{a} e^{-i \omega_{\text{r}}t}-\xi_\text{s})]
\\&+g_3 (\hat{a} e^{-i \omega_{\text{r}}t}-\xi_{\text{s}}+\text{h.c.})^3 \\
    &+g_4 (\hat{a} e^{-i \omega_{\text{r}}t}-\xi_{\text{s}}+\text{h.c.})^4 
\end{split}
\end{equation}

Now, to determine $\xi_{\text{s}}$, we collect the terms with $\hat{a}^\dagger  e^{i \omega_{\rm r}t}$:
\begin{center}
$-\xi_{\text{s}} \omega_{a,0}+i\dot{\xi}_{\text{s}}+2 \text{Re}(e^{i\omega_{\text{s}}t})\epsilon_{\text{s}}=0 $ 
\end{center} 
\begin{equation} \label{eq:1}
\xi_{\text{s}}=-\left (\frac{\epsilon_{\rm s,0}}{\omega_{\text{s}}-\omega_{a,0}}e^{-i\omega_{\text{s}}t}-\frac{\epsilon_{\rm s,0}}{\omega_{\text{s}}+\omega_{a,0}}e^{i\omega_{\text{s}}t}\right )
\end{equation}
Considering that both $\xi_{\text{s}}$ and its complex conjugate will appear inside the nonlinear terms we write $\xi_{\text{s}}+\xi^*_{\text{s}}=\xi_{\text{s},\text{eff}} \: e^{-i\omega_{\text{s}}t}+\text{h.c.}$  
\begin{equation} \label{eq:1}
\xi_{\text{s},\text{eff}} =-\left (\frac{\epsilon_{\rm s,0}}{\omega_{\text{s}}-\omega_{a,0}}-\frac{\epsilon_{\rm s,0}}{\omega_{\text{s}}+\omega_{a,0}}\right ).
\end{equation}
This means the first three lines in (A10) cancel up to some time-varying terms, which average to zero very quickly. Now, to eliminate the time dependence, we choose $\omega_\text{r}$ so that the term containing  $\hat{a}^{\dagger 2}$, $\hat{a}^{ 2}$ (which comes from the third-order non-linearity) does not oscillate with time. This condition can be satisfied if  $\omega_{\text{r}}=\omega_{\text{s}}/2$:

\begin{center}
$3g_3  (\hat{a}^\dagger e^{i \omega_{\text{r}}t})^2  \xi_{\text{s},\text{eff}}\cdot e^{-i\omega_{\text{s}}t}=3g_3 \hat{a}^{ \dagger 2} \xi_{\text{s},\text{eff}}\cdot e^{-i(\omega_{\text{s}}-2\omega_{\text{r}})t}.$
\end{center} 
When this condition is satisfied, we can simplify $\hat{H}_{\text{KC}}$ by neglecting every oscillating term to get:
 \begin{equation}
\hat{H}_{\rm KC}= \Delta_{a,\text{r}}\hat{a}^{ \dagger}\hat{a}-K\hat{a}^{ \dagger 2}\hat{a}^{2}+\epsilon_2 \hat{a}^{ \dagger 2}+\epsilon^*_2 \hat{a}^{ 2}-4K \hat{a}^{ \dagger}\hat{a} |\xi_{\text{s},\text{eff}}|^2
 \end{equation}
The first term is a harmonic term weighted by the detuning ($\Delta_{a,\text{r}}=\omega_{a,0}-\omega_{\rm r}$). The second term is the Kerr non-linearity ($K=-6g_4$). The third term is the two-photon drive ($\epsilon_{2}=3g_3\xi_{\text{s},\text{eff}}$), and the last term is the Stark shift induced by the drive. The coherent states $\ket{\pm\alpha}$  are eigenstates of this Hamiltonian with $\alpha^2=\epsilon_2/K$\cite{Shruti_KerrCat_intro, Goto_KerrCat_intro}. The cat states are the orthogonal even- and odd-parity  states formed by the superposition of the coherent states and form two, degenerate ground states of this Hamiltonian :
\begin{equation}
\ket{C_\alpha^\pm}=\frac{1}{\sqrt{1\pm e^{-2\alpha^2}}}\frac{1}{\sqrt{2}}(\ket{\alpha}\pm \ket{-\alpha} ). 
\end{equation}

\subsection{Cat Quadrature Readout }\label{CQR_sub}

The implementation of the cat quadrature readout is similar  to the implementation of the two-photon drive because it relies on the three-wave mixing in the SNAIL. To derive the Hamiltonian for Cat quadrature readout, we first consider the combined Hamiltonian of the SNAIL and the readout resonator $\hat{H}_b$ with a coupling coefficient $g_c$ much smaller than their detuning $\Delta=\omega_{a,0}-\omega_{b,0}$:

\begin{equation}
\hat{H}_0=\hat{H}_{a}+\hat{H}_{\text{s}}+\hat{H}_{b}+\hat{H}_{\text{c}},
\end{equation}
\begin{equation}
\hat{H}_{b}=\omega_{b,0}\hat{b}^{\dagger}\hat{b},
\end{equation}

\begin{equation}
\hat{H}_{c}=g_{\text{c}} (\hat{a}^{\dagger}+\hat{a})(\hat{b}^{\dagger}+\hat{b})\approx g_{\text{c}} (\hat{a}^{\dagger}\hat{b}+\hat{b}^{\dagger}\hat{a}).
\end{equation}
Now, we can cast this Hamiltonian in a dressed form where the bosonic modes are decoupled  by using the following unitary \cite{Blais_JC}:
\begin{equation}
\hat{U_{\text{c}}}=e^{\lambda(\hat{a}^{\dagger}\hat{b}-\hat{b}^{\dagger}\hat{a})}
\end{equation}
The impact on the bosonic operators  $\hat{a}$ and $\hat{b}$ is:

\begin{equation}
\hat{U}_\text{c}\ \hat  {a} \ \hat{U}^\dagger_{\text{c}}= \cos(\lambda)\hat{a}- \sin(\lambda)\hat{b}\approx\hat{a}-\frac{g_{\text{c}}}{\Delta}\hat{b}
\end{equation}
\begin{equation}
\hat{U}_{\text{c}}\ \hat  {b} \ \hat{U}^\dagger_{\text{c}}=  \cos(\lambda) \hat{b}+ \sin(\lambda)\hat{a}\approx\hat{b}+\frac{g_{\text{c}}}{\Delta}\hat{a}
\end{equation}
Expanding $\hat{H}_{a}+\hat{H}_{b}+\hat{H}_{\text{c}}$ we get:
\begin{equation}
\begin{split}
&\omega_{a,0} \left( \cos(\lambda)\hat{a}^\dagger - \sin(\lambda)\hat{b}^\dagger \right)\left ( \cos(\lambda)\hat{a}- \sin(\lambda)\hat{b}\right)\\&+\omega_{b,0} \left( \cos(\lambda)\hat{b}^\dagger+ \sin(\lambda)\hat{a}^\dagger\right)\left ( \cos(\lambda)\hat{b}+ \sin(\lambda)\hat{a}\right)\\&+g_{\text{c}} \left( \cos(\lambda)\hat{a}^\dagger- \sin(\lambda)\hat{b}^\dagger\right)\left( \cos(\lambda)\hat{b}+ \sin(\lambda)\hat{a}\right)+\text{h.c.}\\&+g_3 \left( cos(\lambda)\hat{a}- \sin(\lambda)\hat{b}+\text{h.c.}\right)^3 \\&+g_4 \left( \cos(\lambda)\hat{a}- \sin(\lambda)\hat{b}+\text{h.c.}\right)^4
\end{split}
\end{equation}
Choosing $\lambda$ such that the coupling between the modes (i.e., the coefficient of $\hat{a}^{\dagger}\hat{b}$) vanishes gives:
\begin{equation}
\lambda=\frac{1}{2} {\rm Arctan}\left(\frac{2g_{\text{c}}}{\Delta}\right)\approx \frac{g_{\text{c}}}{\Delta}.
\end{equation}
The renormalized frequencies are $\omega_a=\frac{1}{2}(\omega_{a,0}+\omega_{b,0}+\sqrt{\Delta^2+4g_{\text{c}}^2})\approx \omega_{a,0}+g_{\text{c}}^2/\Delta$ and $\omega_b=\frac{1}{2}(\omega_{b,0}+\omega_{a,0}-\sqrt{\Delta^2+4g_{\text{c}}^2})\approx \omega_{b,0}-g_{\text{c}}^2/\Delta$. Now, the two modes are decoupled, but the mode $\omega_b$ appears in the nonlinear terms of mode $\omega_a$. So $\hat{H}_{a}+\hat{H}_{b}+\hat{H}_{\text{c}}$ takes the effective form:

\begin{equation}
\omega_{a}\hat{a}^{\dagger}\hat{a}+ \omega_b\hat{b}^{\dagger}\hat{b}+g_3 \left( \hat{a}+\frac{g_{\text{c}}}{\Delta}\hat{b}+\text{h.c.}\right)^3+g_4 \left(\hat{a}+\frac{g_{\text{c}}}{\Delta}\hat{b}+\text{h.c.}\right)^4
\end{equation}
The third term is the key to implementing the cat quadrature readout. Adding another drive of the form $\hat{H}_{\text{CQR}}=2 \text{Re}(e^{i\omega_{\text{CQR}}t})(\epsilon_{\text{CQR},0} \hat{a}^{\dagger}+\epsilon^*_{\text{CQR},0}\hat{a})$ can implement the desired interaction and following the same treatment as in part A1 we get:
\begin{equation}
\begin{split}
 &+g_3 \Bigl(\hat{a}e^{-i \omega_{\text{r}}t}+\frac{g_{\text{c}}}{\Delta}\hat{b}e^{-i \omega_{b}t}-\xi_{\text{CQR},\text{eff}}\: e^{-i\omega_{\text{CQR}}t}-\xi_{\text{s},\text{eff}}  e^{-i\omega_{\text{s}}t}\\&+\text{h.c.} \Bigr) ^3 \\
&+g_4 \Bigl(\hat{a}e^{-i \omega_{\text{r}}t}+\frac{g_{\text{c}}}{\Delta}\hat{b}e^{-i \omega_{b}t}-\xi_{\text{CQR},\text{eff}}\: e^{-i\omega_{\text{CQR}}t}-\xi_{\text{s},\text{eff}}\: e^{-i\omega_{\text{s}}t}
\\&+\text{h.c.}\Bigr)^4
\end{split}
\end{equation}
So  from $g_3$, we can also stabilize the following term in the rotating frame provided that $\omega_{}=\omega_b-\omega_{\rm r} $

\begin{equation}\label{CQR}
6g_3 \frac{g_{\text{c}}}{\Delta}\xi^*_{\text{CQR},\text{eff}}\:\hat{a}^\dagger\hat{b}\: e^{i(\omega_{\text{CQR}}-\omega_b+\omega_{\text{r}})t}+\text{h.c.} 
\end{equation}

From $g_4$, there will be two additional terms corresponding to the stark shift from the CQR drive  and another term corresponding to the cross Kerr between the readout and the SNAILmon:
 \begin{equation}\label{eq:38}
 24 g_4(\hat{a}^\dagger \hat{a}+\frac{g_{\text{c}}^2}{\Delta^2}\hat{b}^\dagger \hat{b}) |\xi_{\text{CQR},\text{eff}}|^2 +24 g_4\frac{g_{\text{c}}^2}{\Delta^2}\hat{a}^\dagger \hat{a}\hat{b}^\dagger \hat{b}
  \end{equation}
\begin{figure}[tp]
    \centering
    \includegraphics[width=8cm]{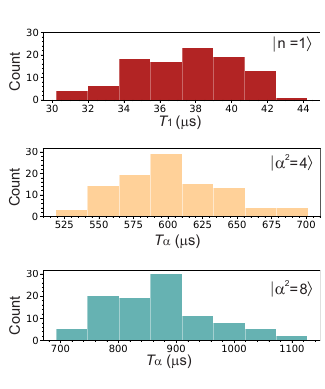}
    \caption{Lifetime statistics. The lifetime for multiple states is examined over the span of two hours with 100 iterations. The lifetime of the first excited state in the Fock basis $\ket{n=1}$ can be used to estimate the single photon loss at the qubit frequency. The long-lived coherent states $\ket{\alpha^2=\epsilon_2/K}$ also fluctuate over time, with an average lifetime of $601$ $  \mu s$ for a cat of size 4 and $872$ $ \mu s$ for a cat of size 8. }
    \label{Fig_statistics}
\end{figure}
\section{Master equation calculations} \label{Master_equation}

In order to simulate the lifetime of the coherent states, we consider the system to be linearly coupled to a macroscopic environment composed of a bath of linear oscillators with continuous modes. The system environment coupling Hamiltonian in the rotating is:
\begin{equation}
\hat{H}_{\rm SB}(t) = \sum_{k}V_{k}(\hat{a}e^{-i\omega_{\rm d}t/2} + \hat{a}^\dagger e^{i\omega_{\rm d}t/2 })(\hat{b}_k(t)+\hat{b}_k^\dagger(t)),
\end{equation}
where $V_k$ gives the coupling strength and $\hat{b}_{k}(\hat{b}_k^\dagger)$ are the annihilation (creation) operators of an excitation of energy $\hbar\omega_k$ in the environment. Following Ref.~\onlinecite{Jaya_SNAIL_theory}, we obtain following master equation up to ${\cal O}(\varphi_{\rm zps}^2)$,
\begin{equation}
\frac{d\hat{\rho}_{\rm S}}{dt} = \frac{i}{\hbar} \left[\hat{H}_{\rm KC},\rho_{\rm S}\right]+\hat{\cal L}_{\rm RWA}\hat{\rho}_{\rm S} + \hat{\cal L}_{\rm NRWA}\hat{\rho}_{\rm S}+ \hat{\cal L}_{\phi}\hat{\rho}_{\rm S},
\end{equation}
where 
\begin{equation}
\hat{\cal L}_{\rm RWA}= \kappa_{\omega_d/2}\bigg[n_{\omega_{\rm d}/2}\mathcal{D}\{\hat{a}^\dagger\}+(1+n_{\omega_{\rm d}/2})\mathcal{D}\{\hat{a}\}\bigg] 
\end{equation}
gives the contributions satisfying the rotating wave approximation. It includes single photon dissipation effects. $\kappa(\omega)$ is the spectral density of the environment defined through $\kappa (\omega) = 2\pi \sum_k |V_k|^2 \delta(\hbar\omega-\hbar\omega_k)$. The magnitude of spectral density depends strongly on the transition energy but is independent of any other system or environment parameters. The thermal population in the environment is given by the Bose-Einstein distribution function, $n(\omega) = \left(e^{\hbar\omega/k_{\rm B}{\cal T}_\omega}-1\right)^{-1}$, where ${\cal T}_\omega$ is the temperature of the environment. Similarly,
\begin{multline}
\hat{\cal L}_{\rm NRWA}= \kappa_{\omega_{\rm d}}\bigg[n_{\omega_{\rm d}}\mathcal{D}\bigg\{\frac{8g_3}{3\omega_{\rm d}}\hat{a}^{\dagger^2}-\bigg(\frac{592g_3}{9\omega_{\rm d}^2}-\frac{16g_4}{g_3\omega_{\rm d}}\bigg)\hat{a}^\dagger\hat{a}\epsilon_2^*\bigg\}\\
+(1+n_{\omega_{\rm d}})\mathcal{D}\bigg\{\frac{8g_3}{3\omega_{\rm d}}\hat{a}^2-\bigg(\frac{592g_3}{9\omega_{\rm d}^2}-\frac{16g_4}{g_3\omega_{\rm d}}\bigg)\hat{a}^\dagger\hat{a}\epsilon_2\bigg\}\bigg] 
\end{multline}
is the beyond-rotating wave approximation contributions. It includes two-photon dissipation effects and drive-induced dephasing, which enters the master equation at ${\cal O}(\varphi_{\rm zps}^{2})$ with $\varphi_{\rm zps}$ being the value of the zero-point fluctuation in the phase operator of the SNAILmon \cite{Jaya_SNAIL_theory}. Finally,
\begin{equation}
\hat{\cal L}_{\phi}= \kappa_{\phi}\mathcal{D}\{\hat{a}^\dagger \hat{a}\} 
\end{equation}
takes into account any additional drive-independent dephasing processes. 

In our model, we consider single-photon effects, multi-photon heating and cooling effects, dephasing as well as non-Markovian detuning noise. The non-Markovian noise is taken to be a Gaussian random fluctuation of the detuning with a mean of $0.03K$ and a standard deviation of $K/500$. The lifetime is averaged over 100 trials of the non-Markovian noise. We estimate $g_3/2\pi = 15 {\rm MHz}$ and the decay rate $\kappa_{\omega_{\rm d}/2}= 1/T_1\approx 26$ kHz from the experiment. We fit the experimental data with ${\cal T}_{\omega_{\rm d}/2}$, ${\cal T}_{\omega_d}$, $\kappa_{\omega_{\rm d}}$, and $\kappa_{\phi}$. For the fit shown in Fig. \ref{fig2}(d), we used $\kappa_\phi=100$ Hz, $\kappa_{\omega_d}=7$ MHz, ${\cal T}_{\omega_d/2}=73.5$ mK, and ${\cal T}_{\omega_d}=515$ mK. We find an excellent agreement between the experimental data and theoretical plot up to the second plateau in the lifetime. The second plateau can be explained by changing the environment temperature (i.e., changing the value of $n_{\rm th}$ to be $5\%$)\cite{frattini2022squeezed}. More statistics on the lifetime time of the Fock qubit and the coherent states are shown in Fig. \ref{Fig_statistics}. Those variations in the lifetime are typical in superconducting qubits \cite{kreikebaum2016optimization, mcewen2022resolving, vepsalainen2020impact, liu2203quasiparticle}.

\section{Universal single-qubit gates for the Kerr cat qubit}
A universal gate set for the Kerr cat qubit has been reported in the literature \cite{First_cat} based on continuous $Z(\theta)$ rotations and a discrete $X(\pi/2)$ rotation on the Bloch sphere. However, the discrete $X(\pi/2)$ relies on turning off the stabilization drive, which protects this qubit and the performance of this gate degrades very quickly with increasing cat size $\left|\alpha\right|^2$\cite{frattini2022squeezed}. Therefore, it is necessary to realize an alternative approach with higher fidelity. Here, we describe the implementation of the gates we introduced in the main text.
\subsection{Continuous Z($\theta$) rotation based on the quantum Zeno effect}

Continuous Z($\theta$) rotations are realized by applying a microwave drive at the frequency $\omega_r = \omega_s/2$, which activates a single-photon transition between the cat states according to:
\begin{equation}
    \hat{H} = \frac{\Omega_{\rm z}}{2}\hat{a}^\dagger + \frac{\Omega_{\rm z}^*}{2}\hat{a}
\end{equation}
where $\Omega_{\rm z}/2$ is the strength of the microwave drive. The strong stabilization drive of the Kerr cat qubit continuously projects its dynamics into the cat space due to the quantum Zeno effects. Therefore, if the single-photon transition rate is much smaller than the energy gap, i.e. $\Omega_{\rm z}\ll E_{\text{gap}}=4\epsilon_2$, one can project the bosonic mode operators into cat space to get:
\begin{equation}
\hat{P}_{\rm C} \hat{a} \hat{P}_{\rm C}=\alpha \hat{Z}-i \alpha e^{-2 \alpha^2} \hat{Y}
\label{eq: proj1}
\end{equation}
\begin{equation}
\label{eq: proj2}
\hat{P}_{\rm C} \hat{a}^\dagger \hat{P}_{\rm C}=\alpha^* \hat{Z}+i \alpha^* e^{-2 \alpha^2} \hat{Y}
\end{equation}
Where $\hat{P}_{\rm C} = |C_\alpha^+\rangle\langle C_\alpha^+|+|C_{\alpha}^-\rangle\langle C_{\alpha}^-|$. We notice that both $\hat{a}$ and $\hat{a}^\dagger$ exponentially approach $\alpha\hat{Z}$ with increasing cat size. When $\Omega_{\rm z}$ is in phase with $\alpha $, the single-photon transition effectively applies continuous $Z(\theta)$ rotation according to the Hamiltonian:
\begin{equation}
    \label{eq: Z}
   \hat{P}_{\rm C} \hat{H}\hat{P}_{\rm C}^\dagger = 2\alpha \Omega_{\rm z}\hat{Z}
\end{equation}\begin{figure}[tp]
    \centering
    \includegraphics[width=8cm]{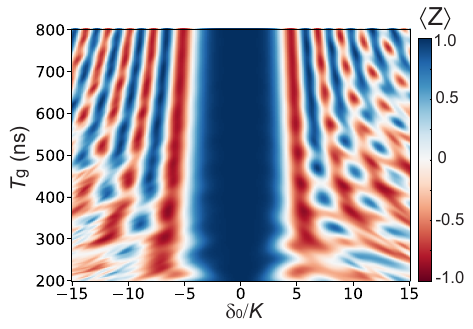}
    \caption{Simulation of the detuning gate. Two $X(\pi/2)$ pulses are used to flip the value of $\left<Z\right>$, similar to the experimental sequence in Fig. \ref{fig3}(b).  Sweeping the gate time $T_g$ and the maximum detuning $\delta_0$ according to Eq.\ref{det_eq} shows multiple regions where the potential deformation leads to coherent tunneling between the energy wells.}
    \label{Fig_xgate}
\end{figure}
 \subsection{$X(\pi/2)$ rotation based on pseudopotential deformation using phase modulation} \label{X gate supp}
 \begin{figure*}[tp]
    \includegraphics[width=17.5 cm]
{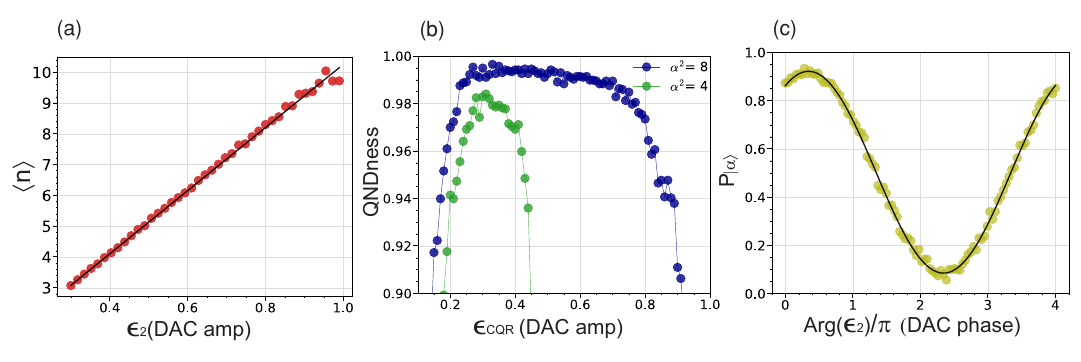}
    \caption{Cat qubit calibrations. (a) The calibration of the cat size as a function of the two-photon drive strength (in units of DAC voltage) is done by measuring the increase in the rabi rate of the cat qubit. The linear trend is expected from the formula $\langle n \rangle =\alpha^2=\epsilon_2/K$ and it reflects the lack of any observable Stark shift from the drive. (b) The cat quadrature readout as a function of the DAC amplitude of the readout tone. We used a 4us readout signal for both cat sizes to maximize the SNR. The readout improves with increasing cat size because we get the bosonic enhancement factor $\alpha$, and for larger cats we can use larger readout signals since the energy gap is proportional to the size (i.e., 4K$\alpha^2$). (c) The mapping of the Fock qubit to the cat qubit enables us to use both encodings if necessary. The pulse sequence is an initial $X(\pi/2)$ pulse in the Fock basis followed by a 2$\mu$s ramp of the stabilization drive and then cat readout. The electrical delay in the input lines determines the phase shift required to prepare the cat qubit in any desired point in the z-y plane. With this protocol, we can prepare the coherent states $\ket{\pm\alpha}$ with a probability $P_\alpha= 93\%$. The data here is fitted to a sinusoidal function.}
    \label{Fig1_supp}
\end{figure*}

Here we describe the Hamiltonian that leads to the novel $X(\pi/2)$ gate introduced in Section \ref{Qubit Control and Readout} of the main text. The $|+Z\rangle$ and $|-Z\rangle$ are approximately two coherent states localized in the two wells of the pseudopotential generated by the Kerr cat Hamiltonian $H_{KC}$ (Fig. \ref{fig1}(b)). Therefore, an X($\pi/2$) gate requires inter-well tunneling to achieve the desired population transfer. By introducing a positive detuning $\Delta\hat{a}^\dagger \hat{a}$ to the Kerr cat Hamiltonian, the energy barrier between the two wells is lowered, allowing inter-well tunneling.
\begin{equation}
\label{eq:KC}
\hat{H}_{\text{KC}}/\hbar=-K \hat{a}^{\dagger2}\hat{a}^2+\epsilon_2\hat{a}^{\dagger2 }+\epsilon^*_2\hat{a}^2 + \Delta\hat{a}^\dagger\hat{a}
\end{equation}
Because $\Delta = \omega_{\rm Q} - \omega_{\rm S}/2$ is the detuning between the qubit frequency and half of the stabilization frequency, there are two ways to introduce such detuning.  First, one can modulate the external flux threading the SNAILs to change the qubit frequency $\omega_Q$. However, this method also changes the Kerr nonlinearity and the SNAIL's third-order nonlinearity, making the analysis more complicated. Therefore, we use the second approach, which involves modulating the stabilization frequency $\omega_S$ directly. Practically, we modulate the phase of the stabilization drive according to:
\begin{equation}
    \epsilon_2(t)=\epsilon_{2}(0) e^{-it\delta_d(t)}
\end{equation}
This time dependence in the drive can be viewed as an effective time-dependent detuning under the following transformation:
\begin{equation}
    \begin{split}
    H_{\text{rot}} &= \hat{U}_{\text{rot}}\hat{H}_{\text{KC}}\hat{U}_\text{rot}^\dagger+i\hbar\dot{\hat{U}}_\text{rot}\hat{U}_\text{rot}^\dagger \\
    &= -K \hat{a}^{\dagger2}\hat{a}^2+\epsilon_2\hat{a}^{\dagger2 }+\epsilon^*_2\hat{a}^2 + \Tilde{\Delta}(t)\hat{a}^{\dagger}\hat{a},
\end{split}
\end{equation}
\begin{equation}
    \hat{U}_\text{rot} = e^{i \delta_{ d} (t) ta^\dagger a/2}
\end{equation}
 The unitary transformation is an effective rotating frame transformation with $\Tilde{\Delta}(t) = -\frac{1}{2}(\delta_d(t) + t \dot{\delta}_d(t))$ as the effective detuning.  The pulse shape we use for the phase modulation is:
\begin{equation}\label{det_eq}
   \delta_d(t)/\delta_0 =
\left\{
	\begin{array}{ll}
		- \sin(\frac{3\pi}{2} t/T_g)  &  t \leq T_g/3 \\
		-\frac{f(t)}{1-f(T_g)} (f(t)-f(T_g)). &  t > T_g/3
	\end{array}
\right.
\end{equation}
The function $f(t)=e^{-\frac{(t-T_g/3)^2}{2\sigma^2}}$ is a Gaussian with a standard deviation $\sigma=T_g/4$. This pulse shape prioritizes the first part of the pulse by making the ramp down much faster than the ramp up to compensate for the term proportional to the derivative. In practice, it offered around a factor of two faster gates compared to a $\sin^2(\pi t/T_g)$ pulse shape. The calibration process for the $X(\pi/2)$ gate is shown in Fig. \ref{fig3}(c); we implement two $X(\pi/2)$  pulses with variable gate time $T_g$ and detuning modulation depth $\delta_0$. The first readout is used to initialize the qubit along the z-axis of the Bloch sphere. The measured expectation value of Pauli Z is shown in Fig. \ref{fig3}(d), reflecting a coherent state population transfer between the $|+Z\rangle$ and $|-Z\rangle$ states. The simulation of this gate is shown in Fig. \ref{Fig_xgate}. For short gate times (i.e., $T_g<300$ ns), the process is not adiabatic with a Kerr of $K/2\pi= 1.2$ MHz.

\section{Kerr-cat calibrations} \label{calibration_supp}
\subsection{Cat Size Calibration}
To calibrate the Kerr-cat qubit we first rely on the Fock qubit calibration of the oscillator frequency $\omega_Q$ and the Kerr nonlinearity $K$. We use standard techniques like two-tone spectroscopy  to measure $\omega_Q$ and estimate the Kerr nonlinearity first by measuring the two-photon transition $K=\omega_{02}/2-\omega_{\rm Q}$ in frequency domain then by using two-tone in pulse fashion in the 1-2 subspace (i.e. using pulse sequence $X_{01}(\pi)X_{12}(\pi,\omega) $ ) and the formula $K=-\alpha/2=-(\omega_{12}-\omega_{01})/2$. 

To calibrate the two-photon drive, we can start by driving the 0-2 transition by driving at $\omega_{20}=2\omega_{\rm Q}-2K$, which leads to a Rabi rate of $\sqrt{8}\epsilon_2$. This method is accurate enough to get the right order of magnitude in the weak limit (i.e., $\epsilon_2 < 2K$). Next, to calibrate the amplitude of the two-photon accurately, we need to observe the change in the Rabi rate of the cat qubit according to $\approx 2\alpha |\Omega_{\rm z}|\cos(\theta_{\rm z})\hat{Z}-2\alpha \Omega_{\rm z} \sin(\theta_{\rm z}) e^{-2|\alpha|^2} \hat{Y}$ which is accurate in the limit $\epsilon_{\rm z} \ll 4K\alpha^2$ and large enough cat size to see the exponential suppression (i.e.,  $\alpha^2\gtrsim  2$). With the definition of the drive as $\frac{\Omega_{\rm z}}{2}(\hat{a}+\hat{a}^\dagger)$we  measure $\Omega_{\rm z}$ directly since it is the Rabi rate of the Fock qubit. However, to measure the cat size, we need to calibrate the phase $\theta_{\rm z}$, which requires creating a chevron similar to Fig. \ref{fig3}(f). The full calibration of the  cat size as a function of the strength of the two-photon drive is shown in Fig. \ref{Fig1_supp} (a), which calibrates the cat size $\alpha^2$ according to 
\begin{equation}\label{cat_size}
    \alpha^2=\left(\frac{\Omega_c}{2\Omega_{\rm z}}\right)^2
\end{equation}
Where $\Omega_c$ is the Rabi rate of the cat qubit.
\subsection{Mapping The Fock qubit to the Cat qubit}\label{mapping}
With the large ratio of the Kerr-nonlinearity relative to the device's single photon decay rate, we can map the Fock qubit to the cat qubit and perform operations in each encoding. By parity conservation, the ground state of the Fock qubit ($\ket{n=0}$) maps to the even cat state, and the excited state  ($\ket{n=1}$) maps to the odd cat state. However, to prepare the coherent states or the imaginary cat states, we can apply an $X_{01}(\pi/2)$ pulse in the Fock basis and then ramp up the stabilization drive, which prepares either the coherent states $\ket{\pm \alpha}$ or the imaginary cat states $\ket{C_{\pm}^i}$ according to the relative phase between the  coherent state $\alpha=(\epsilon_2/K)^{1/2}$ and the initial  $X(\pi/2)$ pulse as shown in Fig. \ref{Fig1_supp}(c). This sequence enables us to use the cat quadrature readout to project the cat qubit along the z-axis, which has better contrast than the dispersive readout in the Fock basis. 
\begin{figure*}[tp]
    \includegraphics[width=17.5 cm]
{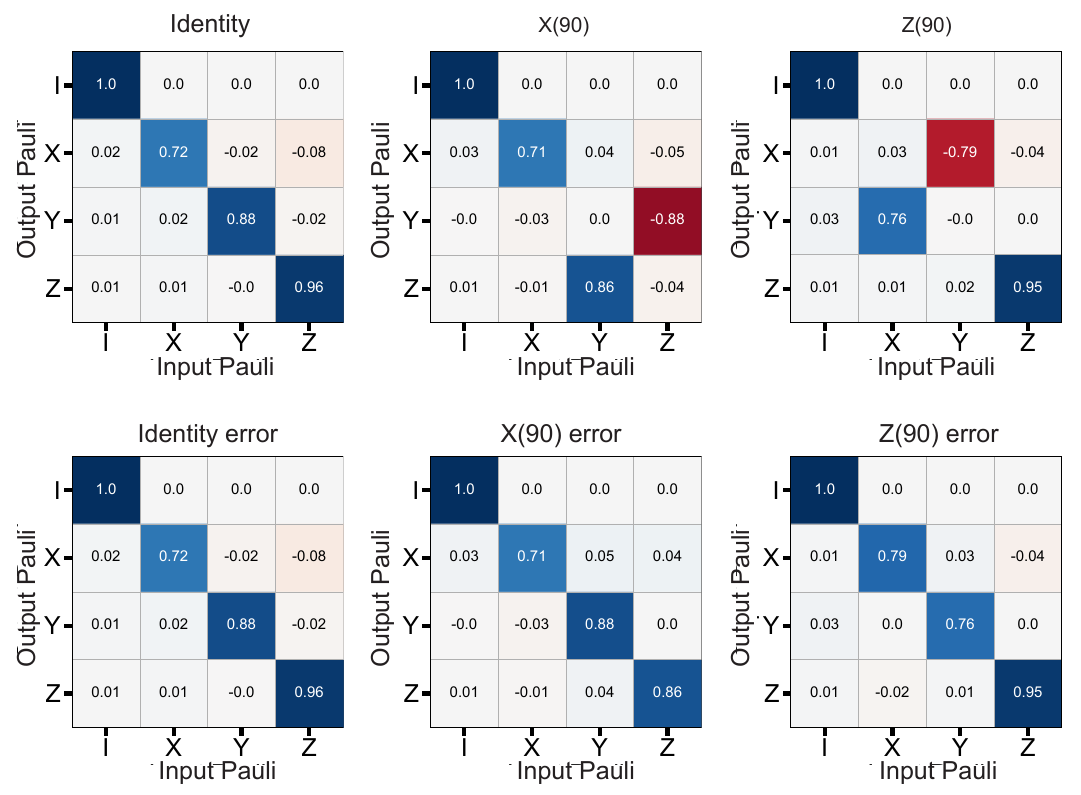}
    \caption{Gates characterization. Pauli transfer matrix for the identity, $X(\pi/2)$ and $Z(\pi/2)$. State preparation and measurement in the X basis is the limiting factor in all three cases. Faster $Z(\pi/2)$ gates can improve this number at the expense of a shorter lifetime for the coherent states.}
    \label{fig_tomog}
\end{figure*}
\subsection{Optimizing the cat quadrature readout}
The theoretical treatment of the cat quadrature readout was introduced in appendix \ref{CQR_sub} and more details can be found in reference \cite{First_cat,frattini2022squeezed}. It involves a microwave-activated beam-splitter interaction between the cat qubit mode $\hat{a} $ and the readout resonator $\hat{b}$. When interaction strength $\epsilon_{\rm CQR}=6g_3\frac{g_c}{\Delta} \xi_{\rm CQR,eff}$ is much smaller than the energy gap $4K\alpha^2$, equation \ref{CQR} can be projected into the computational space giving the effective form:
\begin{equation}
    \hat{H}_{\rm CQR}=\alpha \epsilon_{CQR} \hat{Z} (\hat{b}+\hat{b}^\dagger)
\end{equation}
Which is commonly referred to as longitudinal readout in the field of superconducting qubits  \cite{QuantumEngineerGuide}.  In this readout scheme, a photon from the drive combines with a photon from the qubit through the third-order nonlinearity of the SNAIL to create a photon in the readout mode. The stabilization drive replenishes the qubit much faster than any change in its state, keeping this process quantum non-demolition to first-order.  

To quantify the QNDness, we need two consecutive readouts, as shown in Fig. \ref{fig3}(a), and the results for cat size $\alpha^2 =4,8$ are shown in Fig. \ref{Fig1_supp}(b) as a function of the readout strength. We perform each measurement in $4$ $\mu \rm s$, which is more than two orders of magnitude smaller than the lifetime of the coherent states for both cat sizes.

\section{Gates characterization with process tomography} \label{tomog}
To get an estimate of the quality of the single qubit gates, we use process tomography in the cat encoding and leverage the cat quadrature readout, which we expect to contribute minimally to state preparation and measurement errors. For a cat of size $\alpha^2=4$, the lifetime along the z-axis is around $T_\alpha \approx 600$ $\mu \rm s$, and the lifetime along the X, Y-axes is around $T_C\approx 5$ $\mu \rm s$. We perform the $Z(\pi/2)$ gate in 120 ns and the $X(\pi/2)$ gate in 320 ns. To calculate the gate fidelity, we use $\mathcal{F}_g=\frac{1}{3} (\frac{1}{2}\text{Tr}(\mathcal{R}^T_{\rm ideal} \mathcal{R}_{\rm exp})+1)$, with $\mathcal{R}^T_{\rm ideal} $ being the ideal Pauli transfer matrix (PTM) and $\mathcal{R}_{\rm exp}$ being the experimental value.

To calculate the experimental PTM, we start by using the cat quadrature readout to initialize the qubit in the state $\ket{\alpha}$, then we prepare four different states by applying $I, Y(90), X(270), X(180)$. The final gate is a virtual gate that flips the sign of any subsequent $\Omega_{\rm z}$ drive or $\epsilon_{\rm CQR}$ readout tone. After that, we apply the desired operation and construct the output density matrix for every input state. Finally, we perform  least-square approximation to find the single-qubit PTM. The process matrix for the identity, the $X(\pi/2)$ gate, and the $Z(\pi/2)$ gate from this analysis are shown in Fig. \ref{fig_tomog}. However, this method neglects leakage by enforcing normalization, which can be estimated in future devices with larger Kerr by mapping back to the Fock basis to measure the population of the $\ket{n=3}$ state. Also, in this method, the error is limited by state preparation since  $\mathcal{F}_{X_{\pi/2}}=90.8\%$ and $\mathcal{F}_{Z_{\pi/2}}=91.7\%$, and for the identity, we measure $\mathcal{F}_{I}=92.7\%$. The characterization of those gates in a SPAM-free way that distinguishes the noise bias of every gate in the set is the subject of the next study on this device.
\begin{table}[h!]

\caption{Experimental Parameters}
\centering
\begin{tabular}{ |p{6.6cm}  | c |}
\hline  \hline
 \textbf{Parameter} & \textbf{Value}   \\
\hline    
SNAILmon qubit frequency $\omega_Q/2\pi$& 5.9 GHz     \\ [1ex] 
\hline 
SNAILmon capacitive shunt energy  $E_c/h$ & 118 MHz     \\[1ex]
\hline 
Number of SNAILs& 2      \\ [1ex]
\hline 
SNAILmon junctions asymmetry  & 0.1     \\ [1ex]
\hline 
SNAILmon Kerr nonlinearity $K/2\pi$& 1.2 MHz     \\ [1ex]
\hline 
Fock qubit single-photon decay time $T_1$& 38.5 $\mu$s     \\ [1ex]
\hline 
Fock qubit Ramsey decay time $T_2^*$& 3$\mu$s     \\ [1ex]
\hline \hline  
Readout resonator frequency $\omega_R/2\pi$& 7.1 GHz     \\[1ex]
\hline 
Readout resonator linewidth  $\kappa_R/2\pi$& 0.4 MHz     \\ [1ex]
\hline 
Readout to SNAILmon coupling strength $g/2\pi$& 125 MHz     \\ [1ex]
\hline 
Readout to SNAILmon cross-Kerr  $\chi/2\pi$& 40 KHz     \\ [1ex]
\hline 
Readout resonator internal quality factor $Q_{R,i}$& 3$\times 10^5$     \\ [1ex]
\hline \hline 
Purcell filter frequency $\omega_P/2\pi$& 7.2 GHz     \\ [1ex]
\hline 
Purcell filter linewidth $\kappa_P/2\pi$& 60 MHz     \\ [1ex]
\hline 
\end{tabular}
\label{table:1}
\end{table}
\section{Experimental Parameters}
The experimental parameters of this device were chosen to maximize the lifetime of the qubit and achieve high QND readout while maintaining fast control in the cat subspace. The data are summarized in Table \ref{table:1}. The qubit frequency $\omega_{\rm Q}$ was chosen to be compatible with cryogenic components and room-temperature microwave electronics while maintaining a large ratio of $E_J/E_c$ to maximize the number of bound states in the oscillator ($E_J$ is the effective Josephson energy of the SNAIL). The capacitive shunt energy  $E_{\rm c}/h$ was chosen to be a value between $ 300$ MHz, which is commonly used to make transmon qubits with high coherence and the smaller values previously reported in 3D cat experiments ($E_{\rm c}\approx 64$ MHz), which increases the ratio of the nonlinearities $g_3/g_4$ of the SNAIL. Using two SNAILs instead of one increases the number of bound states by a factor of 4 and increases the ratio of  $g_3/g_4$ by a factor of 2\cite{Nick_SNAIL_Optimization}. The $1.2$ MHz Kerr was an optimal value in practice to maximize the lifetime of the coherent states while maintaining fast single qubit gates. The readout mode linewidth was chosen to be $0.1$-$1$ MHz  to maximize the SNR of the cat quadrature readout \cite{First_cat}. The dedicated, sharp Purcell filter for readout was crucial to maintaining a high limit on the qubit lifetime while keeping the coupling strength between the qubit and readout large (i.e., $g/\Delta \approx 0.1$). The original Purcell limit on the lifetime of the SNAILmon (i.e., $T_1$) from the strongly coupled port is estimated to be  $\approx12$  $\mu$s from finite-element simulation using Ansys High-Frequency Electromagnetic Field Simulation (HFSS). The 30 dB of isolation at the qubit frequency increases this limit to 12 ms.

\section{Band-block filter }\label{filter}

\begin{figure*}[htp]
\includegraphics[width=16 cm]
{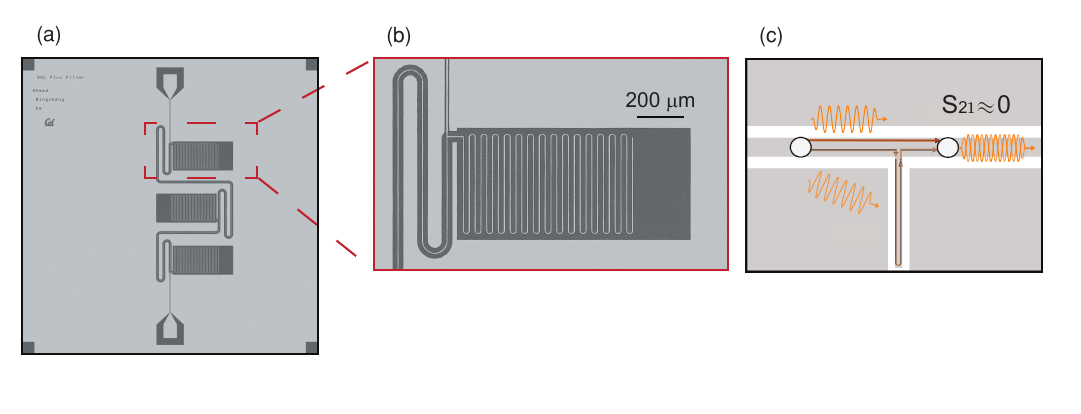}
    \caption{Band block filter. (a) The 5x5 mm$^2$ test device used to measure the scattering parameters in Fig 2. (b) A zoomed view into one stub of the filter where the length is chosen to have destructive interference around the qubit frequency. (c) The destructive interference results from the $\pi$  phase the signal picks from the two paths}
    \label{Fig4_supp}
\end{figure*}

The band-block filter enables strong coupling to the qubit, eliminating the need for very large input power. In the early stages of this experiment, the mixing chamber of the dilution fridge used to heat up by a few mK when we used the control line for pumping. The chip we used to characterize the filter is shown  in Fig. \ref{Fig4_supp} (a) with a zoomed view into one of the stubs in Fig. \ref{Fig4_supp}(b).  The open boundary condition requires this filter design to be $\lambda/4$ to have destructive interference inside the stop band of the filter \cite{Pozar}. Fig. \ref{Fig4_supp}(c)  shows a visualization of this process. The normalized transmission coefficient $\Bar{S}_{21}$ of the filter shown in Fig. \ref{Fig4} includes 69 dB of attenuation in the lines and 10 dB insertion loss just outside the filter range. Adapting this filter design in other experiments involving strong off-resonance drives should make those processes efficient with minimal heating effects. 

\begin{figure*}[btp]
\includegraphics[width=17.5 cm]
{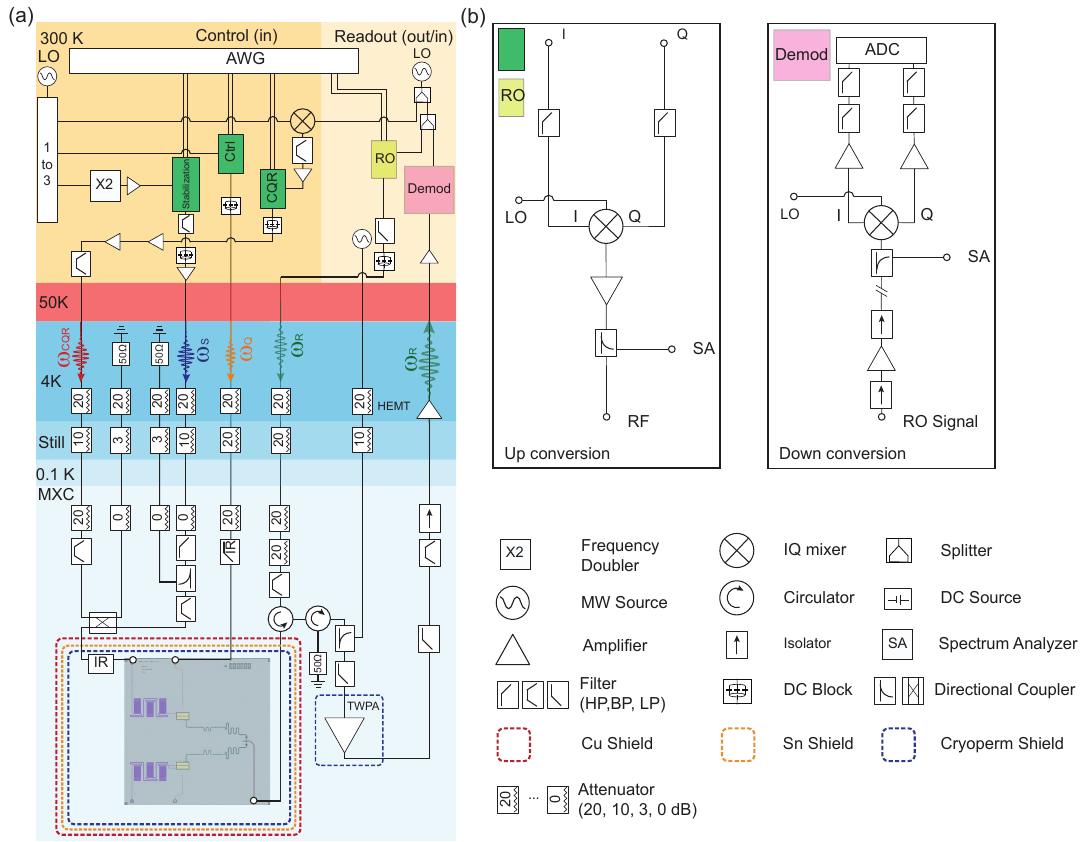}
    \caption{Wiring diagram. (a) The microwave components used in this experiment for control and readout involve an arbitrary waveform generator (AWG), local oscillators (LOs), and an analog-to-digital converter (ADC). The rest of the microwave components are shown. (b) The two schemes of up-conversion and down-conversion of the signals using the AWG and ADC equipment at a 1GS/s sampling rate.}
    \label{Fig_wiring}
\end{figure*}

\section{Microwave control hardware}
As a strongly driven qubit, the Kerr-cat qubit requires multiple, strong drives at different frequency ranges with high spectral purity and fast control. To accomplish this, we use the standard techniques of up-conversion and down-conversion of RF signals that we generate with the arbitrary waveform generator (AWG) and analyze with an analog to digital converter (ADC) at 1 GS/s sampling rate. 

The wiring diagram shown in Fig. \ref{Fig_wiring} includes the different stages of the dilution fridge. The AWG enables using the lower sidebands of the IQ mixer with IF frequencies of 70 MHz for the qubit control signals, 140 MHz for the stabilization drive, and 140 MHz for the cat quadrature readout up-conversion. For the down-conversion of the readout signal, we use 210 MHz IF frequency. The difference between the upconversion and down conversion IF frequencies is the result of the input signal being at $\omega_{\rm CQR}$ and the output signal being at $\omega_{\rm R}$. In this experiment, we use two local oscillators around the qubit frequency and the readout frequency. To generate the carrier of the stabilization drive, we use a frequency doubler, and to generate the carrier of  the cat quadrature readout, we mix the two local oscillators with an RF mixer (shown in the diagram). This arrangement eliminates the sensitivity on the phase of the local oscillators when controlling the qubit (i.e., driving on phase as in Fig. \ref{fig3} (f)) or performing the cat readout (i.e., the IQ signal shown in Fig. \ref{fig3} (b)). Also, since the Kerr cat Hamiltonian is realized in the rotating frame, the readout signal picks a phase depending on the delay we use relative to the time we ramp up the stabilization drive. To account for this delay, we multiply the input signal $\epsilon_{\rm CQR}$ with the appropriate phase to make sure the output signal always aligns with the reference we use to draw the discrimination line in the IQ plane. In this experiment, we ensured spectral purity of $\approx$ 40 dB suppression of the next visible side band, which requires large IF frequencies and the use of both high-pass and low-pass filters to suppress the sidebands and DC bias at each mixer to suppress the carrier. 

To ensure the thermalization of the qubit, we attenuate the lines heavily and use a gold-plated cryogenic attenuator from Quantum Microwave at the mixing chamber in combination with eccosorb filters. To avoid dumping heat at the mixing chamber (MXC), we use directional couplers to attenuate the signal at the MXC plate while directing the signal to the 4 K with 50-ohm termination. The output signal is first amplified with a traveling wave parametric amplifier (TWPA) with a gain of around 20 dB  at the mixing chamber \cite{macklin2015near}, then a HEMT amplifier at 4 K, and a low-noise amplifier at room temperature. The signal is then downconverted with an IQ mixer, filtered around the demodulation frequency,  and amplified before sending it to the ADC.